\documentclass[fleqn,twoside,twocolumn,nofootinbib]{revtex4} 
\usepackage[sec]{ujp} 
\begin{document}
\title[FOUR-FERMION INTERACTION IN THE LOW-ENERGY APPROXIMATION]
{FOUR-FERMION INTERACTION IN THE LOW-ENERGY APPROXIMATION OF QUANTUM CHROMODYNAMICS}%
\author{S.V. Kutnii}
\affiliation{\bitp}
\address{\bitpaddr}
\author{P.I. Golod}%
\affiliation{\bitp}
\address{\bitpaddr}
\udk{539.1.01}  \razd{\seci}

\setcounter{page}{1161}%
\maketitle

\begin{abstract}
A model with the four-fermion interaction is
 derived using the self-consistent field method in the low-energy
  limit of quantum chromodynamics. The resulting Lagrangian contains
  not only the trivial and chiral terms but also the interaction of the
   vector and pseudovector currents.
\end{abstract}

\section{Introduction}

As is known, low-energy effects related  to strong interactions
cannot be described perturbatively. That is why effective models
gain in importance. Models with the four-fermion interaction that
naturally proceed from the calibration field theories are especially
promising, because one can consider them as relativistic analogs of
the Bardeen--Cooper--Schrieffer (BCS) model and apply the methods
successfully used in the theory of superconductivity.

The model developed by Nambu and Jona-Lasinio \cite{NJL} represents
one of the effective theories of this kind. This model was proposed
before the appearance of quantum chromodynamics (QCD) and it does
not contain colors in its original form. It attracted a renewed
interest in the 1980s-1990s. In the
framework of the model, a mass spectrum of mesons was obtained
in good agreement with experiment. A review of the results can
be found in \cite{Volkov1}.

Work \cite{Volkov2} gives the substantiation of the
Nambu--Jona-Lasinio model in the framework of QCD. This
substantiation is based upon the assumption about the structure of
gluon Green functions. The exact form of the gluon propagator is
till now unknown, so this reasoning is not too reliable. That is why
the search for other ways of substantiation is of special
importance.

It turns out that a gluon field, whose Lagrangian includes
nonlinear terms of the third and fourth powers with respect to
the gluon fields, can be considered, by applying the self-consistent
field method (or, what is the same, Bogolyubov’s method of
quasiaverages; see, e.g., \cite{Svidz}). Such an approach was
first proposed by Kondo \cite{Kondo2003, Kondo2004}. Starting from
the Lagrangian for the Yang--Mills free field, Kondo demonstrated the
presence of the gluon condensate in the vacuum state and arrived at
the Lagrangian quadratic in the gluon fields for the mean field
approximation with a mass term caused by the condensate.

The generation of the effective mass of the gluon field results  in
that, in this approximation, its classical potential becomes the
Yukawa one, i.e. quasilocal. That is why, if quarks are added to the
model, the four-fermion interaction of the Nambu--Jona-Lasinio type
will represent the zero-order approximation of the effective quark
Lagrangian.

\section{Self-Consistent Field Method for QCD}

The BCS model in the theory of superconductivity includes
a nonlinearity. For this case, Bogolyubov has developed a special version
of the self-consistent field method. It is important that
Bogolyubov’s method can be reformulated in terms of the continual
integral (see, e.g., \cite{Svidz}), which allows one to apply it
to field theories of various kinds.

Bogolyubov’s approach was applied to the Yang--Mills field by Kondo
(\cite{Kondo2003, Kondo2004}). We will be based on these results.

According to \cite{Kondo2003}, starting from the standard
Yang--Mills Lagrangian and successively introducing the mean fields
$\phi, \varphi^{a}, G_{\mu\nu}, B^{a}_{\mu\nu},$ and $V^{a}_{\mu\nu}$,
one can obtain a Lagrangian that will be only quadratic with respect
to the gluon fields $\mathcal{A}_{\mu}^{a}$:
\begin{equation}
\mathcal{L} = \mathcal{L}_{MF} + \frac{1}{2}\mathcal{A}_{\mu}^{a}
\mathcal{K}^{\mu\nu\,ab}\mathcal{A}_{\nu}^{b} +
\mathcal{A}_{\mu}\mathcal{J}^{\mu}.
\end{equation}

Here, $\mathcal{L}_{MF}$ is the part of the Lagrangian that depends only on the mean fields:
\[
\mathcal{K}_{\mu\nu}^{ab} = \delta^{ab}\left[-(1 - \rho^2)
(\eta_{\mu\nu}\partial^2 - \partial_{\mu}\partial_{\nu}) -
\lambda^{-1}\partial_{\mu}\partial_{\nu}\right]-
\]
\[
- ig\sigma{f}^{abc}B^{c}_{\mu\nu} +
\sigma_{\phi}\delta^{ab}\eta_{\mu\nu}\phi +
\sigma_{\varphi}d^{abc}\eta_{\mu\nu}\varphi^{c}
 +\sigma_{G}\delta^{ab}\times
\]
\begin{equation}
\times\left(G_{\mu\nu} -
\frac{1}{2}\eta_{\mu\nu}G^{\rho}_{\rho}\right)
+\sigma_{V}d^{abc}\left(V^{c}_{\mu\nu} -
\frac{1}{2}\eta_{\mu\nu}V^{c\,\rho}_{\rho}\right).
\end{equation}

The adding of the fermion part describing quarks to the Yang--Mills
Lagrangian will result in the appearance of the quark contribution
to the current $\mathcal{J}^{\mu}$:
\begin{equation}
\mathcal{J}_{q\,\mu}^{a} =
\bar{\psi}_{i}\gamma_{\mu}T^{a}_{ij}\psi_{j}.
\end{equation}

Lagrangian (2) is quadratic with  respect to the gluon fields;
therefore, they can be integrated. The resulting Lagrangian will
include the interaction of the quark currents in the form
\begin{equation}
\mathcal{L}_{qq} =
-\frac{1}{2}\mathcal{J}_{q\,\mu}^{a}\left[\mathcal{K}^{-1}\right]
^{\mu\nu\,ab}\mathcal{J}_{q\,\nu}^{b}.
\end{equation}

In \cite{Kondo2003, Kondo2004}, it was shown that $\phi$ can have a
nonzero value of the vacuum average $\phi_0$. It is obvious that
this vacuum average is connected with the effective gluon mass:
\begin{equation}
|\sigma_{\phi}\phi_0|\sim{m_{\mathcal{A}}^{2}}.
\end{equation}

The numerical value of $m_{\mathcal{A}}$ is of the order of 1 GeV.

With regard for this  fact, $\mathcal{K}^{-1}$ can be expanded in
a series in powers of
$\left(\sigma_{\phi}\phi_0\right)^{-1}$:
\begin{equation}
\left[\mathcal{K}^{-1}\right]^{\mu\nu\,ab} =
\frac{1}{\sigma_{\phi}\phi_0} \eta^{\mu\nu}\delta^{ab} +
O\left(\frac{1}{\sigma_{\phi}^2\phi_0^{2}}\right).
\end{equation}

It is evident that the terms of higher orders  will be small.
Therefore, the expansion will be meaningful if
\begin{equation}
p_{\mu} < \sqrt{\sigma_{\phi}\phi_0},
\end{equation}
and the mean fields will satisfy similar smallness conditions.

In the zeroth order, we obtain the four-fermion interaction
\begin{equation}
\mathcal{L}_{qq}^{(0)}
 = -\frac{1}{2\sigma_{\phi}\phi_0}\bar{\psi}_{i}\gamma_{\mu}T^{a}_{ij}\psi_{j}
 \bar{\psi}_{k}\gamma^{\mu}T^{a}_{kl}\psi_{l}.
 \end{equation}

Condition (7) evidently specifies the low-energy
approximation. The above considerations testify
 to the fact that an adequate description of the low-energy
 approximation of QCD must include a condition of type (7) in
 addition to the model with the four-fermion interaction of the Nambu--Jona-Lasinio type.
  This fact can mean that the momentum cut-off on the scales of the order of
   $m_{\mathcal{A}}$ is the way of regularization to be preferred.
   The problem of choosing the regularization is important, because the Nambu--Jona-Lasinio
    model is not renormalized.

\section{Analysis of Four-fermion Interaction}

Let us write the quark part of the Lagrangian keeping only the zeroth order of expansion (6):
\begin{equation}
\mathcal{L}_q = \bar{\psi}_{i}\left(i\hat{\partial}
 - m_0\right)\psi_i-\frac{1}{2\sigma_{\phi}\phi_0}\bar{\psi}_{i}
 \gamma_{\mu}T^{a}_{ij}\psi_{j}\bar{\psi}_{k}\gamma^{\mu}T^{a}_{kl}\psi_{l}.
 \end{equation}

If $T^a$ are the generators of the Lie algebra
$su(N)$, then the four-fermion term can be simplified.

Let a bilinear form be specified on $gl(N)$:
\begin{equation}
\langle{X,\,Y}\rangle \equiv
 {\rm Tr}\left[XT^aYT^a\right] = X_{ij}T^a_{jk}Y_{kl}T^a_{li}.
 \end{equation}

This form will be invariant with respect to the adjoint action of the $GL(N)$ group:
\[
 \langle{\Omega(X),\,\Omega(Y)}\rangle ={\rm Tr}\left[X\omega^{-1}T^a\omega{Y}\omega^{-1}T^a\omega\right]=
\]
\begin{equation}
= \Omega^{ab}\Omega^{ac}{\rm Tr}\left[XT^bYT^c\right],
 \end{equation}
\[
\Lambda\delta^{ab} = {\rm Tr}\left[T^aT^b\right] = {\rm
Tr}\left[\omega^{-1}T^a\omega\omega^{-1}T^b\omega\right] =
\]
\begin{equation}
= \Omega^{ac}\Omega^{bd}{\rm Tr}\left[T^cT^d\right] =
\Lambda\Omega^{ac}\Omega^{bc} \Rightarrow
\end{equation}
\begin{equation}
\langle{\Omega(X),\,\Omega(Y)}\rangle = \langle{X,\,Y}\rangle .
\end{equation}

The basis of the $T^a$ generators can be chosen always in such a way that relation (12) is satisfied.

It is known (see for example \cite{Gholod}) that, for the Lie
matrix groups, any Ad-invariant bilinear form can be represented as
follows:
\begin{equation}
\langle{X,\,Y}\rangle = \lambda{{\rm Tr}}\left[XY\right] + \mu{{\rm
Tr}}X{\rm Tr}Y.
\end{equation}

It implies that, for form (10),
\begin{equation}
T^a_{ij}T^a_{kl} = \lambda\delta_{ij}\delta_{kl} +
\mu\delta_{jk}\delta_{il}.
\end{equation}

Thus, the Lagrangian can be converted to the form
\[
\mathcal{L}_q = \bar{\psi}_{i}\left(i\hat{\partial} -
m_0\right)\psi_i
 -  \frac{1}{2\sigma_{\phi}\phi_0}\biggl[ \lambda\bar{\psi}_{i}\gamma_{\mu}
 \psi_{i}\bar{\psi}_{k}\gamma^{\mu}\psi_{k} +
 \]
\begin{equation}
 + \mu\bar{\psi}_{i}\gamma_{\mu}\psi_{k}\bar{\psi}_{k}\gamma^{\mu}\psi_{i}\biggr].
\end{equation}

As was already noted, the Nambu--Jona-Lasinio model contains no
colors. This fact has a physical sense, because we know that
particles with free color charges are not observed at low energies.
However, Lagrangian (16) includes the last term that is not diagonal, at first
sight, with respect to colors as the quadratic or
the first four-fermion ones. Applying the mean field method to the
Lagrangian written in such a form (by analogy with \cite{NJL}), one would
have to introduce the mean field with free color indices violating
the global color symmetry, which would be nonphysical, because all
particles observed at low energies are ``white'', i.e. have no free
color charges. However, it turns out that the last term is actually
equivalent to the sum of the terms diagonal with respect to colors
similarly to the first four-fermion term.

According to the Fierz theorem \cite{Itzyckson},
\[
\mathcal{L}_q = \bar{\psi}_{i}\left(i\hat{\partial} -
m_0\right)\psi_i - \frac{1}{2\sigma_{\phi}\phi_0}\times
\]
\[
\times\biggr[\mu\bar{\psi}_i\psi_i\bar{\psi}_k\psi_k -
\mu\bar{\psi}_i\gamma^5\psi_i\bar{\psi}_k\gamma^5\psi_k +
\left(\lambda - \frac{\mu}{2}\right)\times
\]
\begin{equation}
\times
\bar{\psi}_{i}\gamma_{\mu}\psi_{i}\bar{\psi}_{k}\gamma^{\mu}\psi_{k}
-\frac{\mu}{2}\bar{\psi}_{i}\gamma^5\gamma_{\mu}\psi_{i}\bar{\psi}_{k}\gamma^5\gamma^{\mu}\psi_{k}\biggl].
\end{equation}

Let us find $\lambda$ and $\mu$. For this purpose, we use such a normalization of the generators that
\[
 T^aT^b = \frac{1}{2N}\delta^{ab} + \frac{1}{2}\left(if^{abc} +
 d^{abc}\right)T^c,
 \]
 \begin{equation}
f^{ace}f^{bce} = {N}\delta^{ab}.
\end{equation}

Then we obtain
\[
 \langle{I, I}\rangle =  {\rm Tr}\left[\sum\limits_{a}\left(T^a\right)^2\right]=
 \frac{N^2 - 1}{2} = \lambda{N} + \mu{N^2},
\]
\[
 \langle{T^a, T^b}\rangle = {\rm Tr}\left[T^aT^cT^bT^c\right] =
\]
\[
 ={\rm Tr}\left[T^aT^b\sum\limits_c\left(T^c\right)^2\right] +if^{ace}{\rm Tr}\left[T^bT^cT^e\right]
 =
\]
\[
= \frac{N^2 - 1}{2N}{\rm Tr}\left[T^aT^b\right]
 -\frac{1}{2}f^{ace}f^{cef}{\rm Tr}\left[T^bT^f\right] =
\]
\begin{equation}
= \frac{N^2 - 1}{4N}\delta^{ab}  -
  \frac{N}{4}\delta^{af}\delta^{bf}=
  \frac{\lambda}{2}\delta^{ab}.
\end{equation}

Thus,
\begin{equation}
\lambda = -\frac{3}{2N},~~~\mu=\frac{1}{2} + \frac{1}{N^2}.
\end{equation}

The four-fermion Lagrangian takes the form
\[
\mathcal{L}_q = \bar{\psi}_{i}\left(i\hat{\partial} -
m_0\right)\psi_i-\frac{1}{4\sigma_{\phi}\phi_0}\times
\]
\[
 \times\biggl[\bar{\psi}_i\psi_i\bar{\psi}_k\psi_k -
\bar{\psi}_i\gamma^5\psi_i\bar{\psi}_k\gamma^5\psi_k -
 \frac{N+2}{2N}\times
\]
\begin{equation}
\times\bar{\psi}_{i}\gamma_{\mu}\psi_{i}\bar{\psi}_{k}\gamma^{\mu}\psi_{k}-\frac{1}{2}\bar{\psi}_{i}\gamma^5\gamma_{\mu}\psi_{i}\bar{\psi}_{k}\gamma^5\gamma^{\mu}\psi_{k}\biggr].
\end{equation}

\section{Conclusions}

The application of the mean field  method to the gluon dynamics
allows one to obtain a model with the four-fermion interaction as a
controlled approximation in the low-energy limit of quantum
chromodynamics. The structure of the effective Lagrangian is
generally similar to that obtained in \cite{Volkov2}, so our
consideration can be treated as a substantiation of the assumptions
about the structure of gluon Green functions.

It is important that the structure of the Lagrangian  enables one to
introduce colorless quark condensates. Their consideration will be
the subject of the following work.

\begin{flushright}
{\footnotesize Received 02.04.09.\\ Translated from Ukrainian by H.G. Kalyuzhna}
\end{flushright}

\rezume{%
ЧОТИРИФЕРМІОННА\\ ВЗАЄМОДІЯ У НИЗЬКОЕНЕРГЕТИЧНОМУ\\ НАБЛИЖЕННІ
КВАНТОВОЇ ХРОМОДИНАМІКИ}{С.В. Кутній, П.І. Голод} {Здійснено
послідовне виведення моделі з чотириферміонною взаємодією
 у низькоенергетичному наближенні КХД методом самоузгодженого поля.
 Результуючий лагранжіан, крім тривіального та кірального доданків,
  містить взаємодію векторних та псевдовекторних струмів.}

\end{document}